\documentstyle[aps,prb,eqsecnum,preprint,tighten]{revtex}
\title{Binary neutron star inspiral, LIGO, and cosmology\footnote{This
work was supported grants from the Alfred P.~Sloan Foundation and
the National Science Foundation (PHY~9308728).}}
\author{Lee Samuel Finn}
\address{Department of Physics and Astronomy,\\
Northwestern University, Evanston, Illinois 60208-3112}
\date{\today}
\begin{document}
\maketitle
\section{Introduction}

The most promising anticipated source for the United States Laser
Interferometer Gravitational-wave Observatory (LIGO)
\cite{abramovici92a} or its French/Italian counterpart VIRGO
\cite{bradaschia90a} is the radiation emitted during the final
moments of inspiral before the coalescence of a neutron star - neutron
star (ns-ns) binary system. The instruments that will operate in the
LIGO facility will evolve over time, and the often-discussed
``advanced detectors'' will be sensitive to inspirals at luminosity
distances greater than\cite{finn93a} 1~Gpc.  At such great distances
the rate and character of the inspiral signal depends on the Hubble
expansion, and several authors have investigated how LIGO and/or VIRGO
binary inspiral observations can be used to measure the Hubble
constant and other cosmological
parameters\cite{schutz86a,chernoff93a,markovic93a}.  This brief report
summarizes a more lengthy investigation into estimates of the rate and
character of the inspiral signals that LIGO and VIRGO should
observe. A complete presentation, which also discusses black hole
binaries, is in preparation for submission to Physical Review D.

\section{Detection by LIGO and VIRGO}


To speak of detection rates implies the choice of a detection
criterion. Here I adopt a particularly simple criterion: an
inspiraling binary is said to have been detected if the anticipated
signal-to-noise ratio $\rho$ in the detector is greater than a fixed
threshold $\rho_0$.

Detector noise may conspire to appear as an inspiraling binary,
leading to a false alarm; or, noise may mask the signal from an
inspiraling binary, leading to false dismissal. The importance of
false alarms increases as $\rho_0$ is decreased, while the importance
of false dismissals increases as $\rho_0$ is increased. I will not
consider either false alarms or false dismissals here.  Preliminary
estimates indicate that a threshold of $\rho_0=8$ corresponds to a
``false alarm'' rate of less than $1\,\text{yr}^{-1}$ for binary ns
inspiral\cite{schutz91a}.


The ns-ns binary inspiral detection rate will be proportional to the
ns-ns binary coalescence rate density. The best estimate of this rate
density at the current epoch is \cite{narayan91a,phinney91a}
$\dot{n}_0=1.1\times10^{-7}h\,\text{Mpc}^{-3}\;\text{yr}^{-1}$, where
$h$ is the Hubble constant in units of
$100\,\text{Km}\,\text{s}^{-1}\,\text{Mpc}^{-1}$.  Estimates of
$\dot{n}_0$ rely on the 3 observed binary pulsar systems that will
coalesce in less than a Hubble time. Phinney has estimated that, while
unlikely, the actual rate could be two orders of magnitude higher or
lower without coming into serious conflict with current
observations\cite{phinney91a}.


Given the detection criterion $\rho>\rho_0$, not all binaries within a
fixed distance will be detected:
\begin{trivlist}
\item[\indent1.]{\em More massive binaries are visible at greater
distances than less massive one.}  The signal-to-noise ratio depends
directly on the intrinsic chirp mass ${\cal M}_0$, where
\begin{equation}
{\cal M}_0\equiv\mu^{3/5}M^{2/5}
\end{equation}
and $\mu$ and $M$ are the reduced and total mass of the binary.
\item[\indent2.] {\em Binary systems are not isotropic radiators.} The
intensity of the radiation in each polarization mode depends on
the orientation of the binary system with respect to the line of sight
to the detector.
\item[\indent3.] {\em The detector antenna pattern is not isotropic.}
LIGO and VIRGO are more sensitive to binaries in one part of the sky
than binaries in another.
\end{trivlist}

When all these effects are taken into account, but cosmological
ones are ignored, the expected rate of detection of ns-ns binary
inspiral in the advanced LIGO detector is found to be\cite{finn93a}
\begin{equation}
{dN\over dt d{\cal M}_0}(\rho>\rho_0) =
69.\,\text{yr}^{-1}\;
{\dot{n}_0\over8\times10^{-8}\,\text{Mpc}^{-3}\,\text{yr}^{-1}}
\left(8\over\rho_0\right)^3
\left({\cal M}_0\over1.2\,\text{M}_\odot\right)^{5/2} P({\cal M}_0),
\end{equation}
where $P({\cal M}_0)$ is the probability that a ns-ns binary has
{\em intrinsic chirp mass} ${\cal M}_0$. Observational\cite{finn94b} and
theoretical\cite{woosley92a} evidence suggests that the ns
distribution is sharply peaked about $1.4\,\text{M}_\odot$,
corresponding to an intrinsic chirp mass of $1.2\,\text{M}_\odot$. If
the mass of all neutron stars is assumed to be $1.4\,\text{M}_\odot$,
then $P({\cal M}_0)$ is a delta-function.


The Hubble expansion modifies these results in several ways:
\begin{trivlist}
\item[\indent1.]{\em Distant sources are more luminous.} The
signal-to-noise ratio of an inspiral event depends directly on the
binary system's chirp mass
\begin{equation}
{\cal M} \equiv {\cal M}_0(1+z),
\end{equation}
where $z$ is the binary system's cosmological redshift; thus, the
Hubble expansion leads to an {\em increase} in the visibility of
moderately distant sources relative to what would be expected without
accounting for the redshift. Note that this is in the opposite sense
to the usual cosmological $K$-correction: ns-ns binary inspiral is
more luminous at higher frequencies than at lower ones.
\item[\indent2.] {\em Distant sources are less frequent.} The cosmological
expansion redshifts the rate of distant coalescence events, which
reduces their contribution to the {\em observed} inspiral rate.
\item[\indent3.] {\em Cosmological volumes are not proportional to the cube of
the luminosity distance.} The volume of spheres of constant luminosity
distance $d_L$ increases more slowly than $d_L^3$; consequently, the
number of sources at great distances is not as large as would be
expected if cosmological effects could be neglected.
\item[\indent4.] {\em Distant sources are older.} The number density of
coalescing binaries and the mass distribution of their components
changes as the universe evolves. Estimates of the density and
distribution in ${\cal M}_0$ at the present cosmological epoch do not
necessarily apply at earlier times.
\item[\indent5.] {\em ${\cal M}$ --- not ${\cal M}_0$ --- is observed.} When a
binary system is observed, ${\cal M}$ is measured. Since ${\cal M}$
depends on $1+z$ the observed binaries will be distributed across a
range of ${\cal M}$. The nature of the distribution depends on the
Hubble constant, which permits the observed distribution to be used to
measure $H_0$\cite{chernoff93a,finn95up}.
\end{trivlist}

Assuming an Einstein-deSitter cosmology, negligible evolution over the
observed sample of neutron stars, and that the intrinsic chirp mass of
all ns binaries is a constant ${\cal M}_0$, the anticipated rate of
detected ns-ns inspiral can be written as an expansion in large
$\rho_0$:
\begin{eqnarray}
{dN\over dt}\left(\rho>\rho_0\right) &=& 69.
{\dot{n}_0\over 8\times10^{-8}\,\text{Mpc}^{-3}\;\text{yr}^{-1}}
\left(8\over\rho_0\right)^3
\left({\cal M}_0\over1.2\text{M}_\odot\right)^{5/2}
\nonumber\\
&&\qquad{}\times
\left[
1 -
  0.32 x
+ 0.05 x^2
+ {\cal O}\left(x^3\right)
\right]\label{eqn:dndM0-approx}
\end{eqnarray}
where
\begin{equation}
x \equiv {8 h\over\rho_0}
\left({\cal M}_0\over1.2\,\text{M}_\odot\right)^{5/6}. \label{eqn:x}
\end{equation}
Assuming the ``best guess''\cite{phinney91a} $\dot{n}_0$ and ${\cal
M}_0$ corresponding to ns-ns binaries composed of
$1.4\,\text{M}_\odot$ neutron stars, $dN/dt$ ranges between 29 and
$43\,\text{yr}^{-1}$ as $h$ ranges from 0.5 and 0.8.

There is a maximum distance (or redshift) beyond which a ns-ns binary
of fixed ${\cal M}_0$ will always have a signal-to-noise ratio less
than $\rho_0$; for ns-ns binaries this limiting redshift is given
approximately by
\begin{equation}
z_0\simeq0.473 x \left[1 + 0.276 x + 0.056 x^2 + {\cal O}(x^3)\right].
\label{eqn:z0}
\end{equation}
Assuming ${\cal M}_0$ corresponding to ns-ns binaries composed of
$1.4\,\text{M}_\odot$ neutron stars, $z_0$ ranges between 0.27 and
0.48 as ranges $h$ from 0.5 and 0.8.

Since $\cal M$ depends on $z$, the distribution of observed chirp
masses will reflect how the spatial volume increases with $z$. The
signal-to-noise ratio $\rho$ depends on the luminosity distance $d_L$;
consequently, the number of binaries observed with $\rho>\rho_0$
depends on how the volume of space increases with luminosity distance.
Thus, the observed distribution of $\cal M$ in binaries with
$\rho>\rho_0$ depends on the relationship between redshift and
luminosity distance, and can be used to measure the Hubble constant.
This is the crux of the cosmology test first proposed by Chernoff and
Finn\cite{chernoff93a}. Preliminary estimates show that observation of
200 ns-ns binaries with $\rho>8$ are sufficient to determine $h$ to
within 10\%.

\section{Conclusions}

Given present estimates of the ns-ns binary coalescence rate and the
ns mass distribution the advanced LIGO detectors can expect to observe
between 30 (if $h=0.5$) and $40\,\text{yr}^{-1}$ (if $h=0.8$) ns-ns
binary inspirals with signal-to-noise ratio $\rho$ greater than
$\rho_0=8$. Present estimates of the coalescence rate are uncertain to
several orders of magnitude, and the actual rate could be two orders
of magnitude higher. The observed binaries will be at redshifts less
than 0.48 in a $h=0.8$ universe, or 0.27 in a $h=0.5$ universe. The
observed distribution of binary chirp masses reflects $h$; under these
conditions, 200 observations are sufficient to determine the Hubble
constant to 10\%.


I am glad to thank Kip Thorne for helpful conversations.

\end{document}